\documentclass[11pt,a4paper]{article}
\pdfoutput=1
\usepackage{jheppub}

\usepackage{graphicx}

\usepackage{amssymb}
\usepackage{amsmath}
\usepackage{mathrsfs}
\usepackage{dsfont}

\usepackage{color}
\newcommand{\Eq}[1]{Eq.~(\ref{#1})}

\newcommand{\tab}[1]{Table~\ref{#1}}

\usepackage{amsmath}
\usepackage{slashed}
\usepackage{amscd}
\usepackage{amssymb}
\usepackage{latexsym}
\usepackage{verbatim}
\usepackage{longtable}
\usepackage{mathbbol}
\usepackage{graphicx}
\usepackage{color}
\usepackage{hyperref}
\usepackage[english]{babel}
\newcommand{\Tr}[1]{\mathrm{Tr}\left(#1\right)}
\newcommand{\tr}[1]{\mathrm{tr}\left[#1\right]}

\newcommand{\brackets}[1]{\langle #1 \rangle}
\newcommand{\rbrackets}[1]{\left( #1 \right)}
\newcommand{\sbrackets}[1]{\left[ #1 \right]}

\newcommand{\Real}[1]{\mathrm{Re}\left( #1 \right)}
\newcommand{\Imag}[1]{\mathrm{Im}\left( #1 \right)}
\newcommand{\epow}[1]{\mathrm{e}^{#1}}

\newcommand{\hmu}{\hat{\mu}}
\newcommand{\hnu}{\hat{\nu}}

\newcommand{\Qhat}{\hat{Q}}

\newcommand{\nablal}{\overset{\leftarrow}{\nabla}}
\newcommand{\nablar}{\overset{\rightarrow}{\nabla}}
\newcommand{\nablasl}{\overset{\leftarrow}{\slashed{\nabla}}}
\newcommand{\nablasr}{\overset{\rightarrow}{\slashed{\nabla}}}

\newcommand{\tauone}{\tau^{1}}
\newcommand{\tautwo}{\tau^{2}}
\newcommand{\tauthree}{\tau^{3}}

\newcommand{\gammafive}{\gamma_{5}}

\newcommand{\Pimn}{\Pi_{\mu\nu}}

\newcommand{\fbar}{\bar{f}}

\newcommand{\chibar}{\bar{\chi}}

\newcommand{\balign}{\begin{align}}
\newcommand{\ealign}{\end{align}}
\newcommand{\beq}{\begin{equation}}
\newcommand{\eeq}{\end{equation}}
\newcommand{\balignat}[1]{\begin{alignat}{#1}}
\newcommand{\ealignat}{\end{alignat}}

\newcommand{\bfig}{\begin{figure}}
\newcommand{\efig}{\end{figure}}

\newcommand{\bc}{\begin{center}}
\newcommand{\ec}{\end{center}}

\newcommand{\btab}{\begin{table}}
\newcommand{\etab}{\end{table}}

\newcommand{\bcom}{\begin{comment}}
\newcommand{\ecom}{\end{comment}}

\newcommand{\bitem}{\begin{itemize}}
\newcommand{\eitem}{\end{itemize}}

\newcommand{\benum}{\begin{enumerate}}
\newcommand{\eenum}{\end{enumerate}}

\newcommand{\xvec}{\vec{x}}

\newcommand{\partialbar}{\bar{\partial}}

\newcommand{\deltamunu}{\delta_{\mu\nu}}
\newcommand{\munu}{{\mu\nu}}

\newcommand{\Llagrange}{\mathcal{L}}

\newcommand{\order}[1]{\mathcal{O}\left(#1\right)}

\newcommand{\ronetwofive}{\mathcal{R}^{1,2}_5}

\newcommand{\refeq}[1]{(\ref{#1})}

\newcommand{\reftab}[1]{\{\ref{#1}\}}

\begin{document}

% \preprint{DESY 14-221}
% \preprint{SFB/CPP-14-93}
% \preprint{HU-EP-14/55}

\title{The hadronic vacuum polarization and automatic $\order{a}$ improvement for twisted mass 
fermions}

\author[a]{Florian Burger,}
\author[a]{Grit Hotzel,}
\author[b]{Karl Jansen,}
\author[c]{Marcus Petschlies}

\affiliation[a]{Humboldt-Universit\"at zu Berlin, Institut f\"ur Physik, Newtonstr. 15, D-12489 Berlin, Germany }
\affiliation[b]{NIC, DESY, Platanenallee 6, D-15738 Zeuthen, Germany }
\affiliation[c]{The Cyprus Institute, P.O.Box 27456, 1645 Nicosia, Cyprus}

\emailAdd{florian.burger@physik.hu-berlin.de, grit.hotzel@physik.hu-berlin.de, karl.jansen@desy.de, m.petschlies@cyi.ac.cy}


\abstract{
The vacuum polarization tensor and the corresponding vacuum polarization
function are the basis for calculations of numerous 
observables in lattice QCD. Examples are the hadronic contributions to 
lepton anomalous magnetic moments, the running of the electroweak and strong couplings and quark masses. 
Quantities which are derived from the vacuum polarization tensor  
often involve a summation of current correlators over all distances in position space leading 
thus to the appearance of short-distance terms.
The mechanism of $\order{a}$ improvement in the
presence of such short-distance terms is not directly covered by the usual 
arguments of on-shell improvement of the action and the operators for a given quantity. If such short-distance contributions appear,
the property of $\order{a}$ improvement needs 
to be reconsidered.  
We discuss the effects of these short-distance terms on the vacuum polarization function for twisted mass lattice QCD   
and find that even in the presence of such terms  
automatic $\order{a}$
improvement is retained if the theory is tuned to maximal twist.
}

\arxivnumber{1412.0546}

\maketitle

\section{Introduction}
\label{sec:intro}
The computation of hadronic contributions to observables derived from the vacuum polarization function, especially the muon anomalous
magnetic moment, $a_{\mu}^{\rm had}$, have recently been a major target of the lattice community, 
see for instance~\cite{Blum:2002ii, Gockeler:2003cw, Aubin:2006xv,
Feng:2011zk, Boyle:2011hu, DellaMorte:2011aa, Bernecker:2011gh, Aubin:2012me, Renner:2012fa, Feng:2013xsa, Burger:2013jya, 
Blum:2013xva}.
The reason is that $a_{\mu}^{\rm had}$ is a prime candidate to find 
indications of physics beyond the standard model. 
One basic element to obtain the leading contribution to $a_{\mu}^{\rm had}$ and other quantities derived from the 
vacuum polarization function with good accuracy 
is the property of 
$\order{a}$
improvement, which guarantees that physical quantities 
scale with a rate of $\order{a^2}$ towards the continuum limit. 

For twisted mass fermions at maximal twist automatic $\order{a}$ improvement has been established for
physical quantities without short-distance singularities~\cite{Frezzotti:2003ni} based on symmetry arguments 
of the lattice theory, see also~\cite{Shindler:2007vp} for a review. 
The hadronic vacuum polarization function in momentum space, $\Pimn(Q)$,
however, also receives short-distance contributions
arising from the Fourier summation of the 2-point vector current correlator $\brackets{J_\mu(x)\,J_\nu(y)}$ 
for $x - y \to 0$. 

Employing Symanzik's effective theory~\cite{Symanzik:1983dc, Symanzik:1983gh}, 
we show in the following that with our
definition of the hadronic vacuum polarization function and at maximal twist
these short-distance contributions do not spoil the
automatic $\order{a}$-improvement of the vacuum polarization function in the
twisted mass formulation of lattice QCD (tmLQCD). 
This finding is in accordance with a similar analysis performed for 
the chiral condensate 
and the topological susceptibility 
\cite{Cichy:2013yea, Cichy:2013zea,Cichy:2014yca}, which also involve summations over all lattice points 
with the corresponding appearance of short-distance contributions.
 
To demonstrate the $\order{a}$ improvement of the complete vacuum polarization 
function we will perform an operator product expansion (OPE)
and determine the operators appearing at small distances. 
An essential step is to 
also identify all operators that can mix with the ones in the OPE. 
This 
requires the investigation of the symmetry properties of all operators 
of equal and lower dimension. The classification of such operators up to 
mass 
dimension 6 are compiled in appendix \ref{app:3}. This classification
can also be useful to 
identify the renormalization pattern of other operators built from 
twisted mass fermions.

The article is structured as follows. 
In Sect.~\ref{sec:def} we state the momentum space definition of the hadronic vacuum polarization 
function whose short-distance contributions we investigate later on. We then briefly outline our strategy to prove 
automatic $\mathcal{O}(a)$ improvement in Sect.~\ref{sec:procedure}. In Sect.~\ref{sec:symmetry_projection} and 
appendix~\ref{app:spacetime_proj} we discuss the position 
space properties of the definition given in Sect.~\ref{sec:def}. 
Sect.~\ref{sec:mixing_of_polarization_tensor} and the appendices 
\ref{app:2} and \ref{app:3} contain a list of the symmetries of the 
twisted mass lattice action and the corresponding classification of 
possible mixing operators. The Symanzik expansion constructed from these operators for the 
vacuum polarization function  
of the local vector current is presented in Sect.~\ref{sec:Symanzik_expansion_local_case}. In 
Sect.~\ref{sec:application_to_conserved_current_correlator} the discussion is 
extended to the case of the conserved vector current. Our 
conclusions follow in Sect.~\ref{sec:conclusions}. 

\section{Definition of the vacuum polarization function}
 \label{sec:def}
To keep the paper self-contained, we give here the expressions of the fermion actions used in
our lattice calculation of the muon anomalous magnetic moment~\cite{Burger:2013jya} in the twisted basis
for a setup of active, mass-degenerate up and down and non-degenerate strange and charm quarks ($N_f=2+1+1$). We will
restrict the discussion to the valence quark sector. 
For details about the sea sector and the simulation setup for $N_f = 2+1+1$  twisted 
mass lattice QCD we refer
to~\cite{Baron:2010bv, Baron:2010th}.

In the valence sector we formally introduce three doublets of quarks: the light quark pair  $\chi_l = (\chi_l^+,\,\chi_l^-) = (u,\,d)$,
a strange quark pair $\chi_s = (\chi_s^+,\,\chi_s^-)$  and a charm quark pair $\chi_c = (\chi_c^+,\,\chi_c^-)$. The superscript sign refers to the
sign of the twisted quark mass for  the corresponding field in the valence Dirac operator. Since we employ the Osterwalder-Seiler
action~\cite{Osterwalder:1977pc,Frezzotti:2004wz} in the heavy sector, the complete valence
action can be written concisely as a sum over standard
twisted  mass action terms for the fermion doublets~\cite{Frezzotti:2003ni},
\begin{equation}   
S_F^\mathrm{val} =\sum\limits_{q=l,s,c}\,\sum\limits_x \chibar_q(x) \left[ D_W +i \mu_q \gamma_5 \tau^3 \right] \chi_q(x)\,.
 \label{eq:valence_action} 
\end{equation}
$\mu_q$
denotes the bare twisted quark mass for flavor pair $q$ (taken positive) and $\tauthree$ is the third Pauli matrix acting in the flavor (sub-)space
spanned by the quark doublet $\chi_q$. Besides $S_F^\mathrm{val}$ we assume as usual an action term for the ghost fields corresponding to
the valence sector. Contact to the physical quark content is made by identifying $(u,d) \leftrightarrow (\chi_l^+, \chi_l^-)$,  $s
\leftrightarrow \chi_s^-$ and $c \leftrightarrow \chi_c^+$. We thus choose these fields to initially construct the  electromagnetic current operator
as a Noether current resulting from the infinitesimal vector variation 
\begin{align*}  
\delta_V \chi = i \alpha(x) \,Q_{\rm em}\,\chi(x)\\   \delta_V
\chibar= - i \alpha(x)\,\chibar(x)\,Q_{\rm em}\,,
 \end{align*} 
with $ Q_{\rm em}= \mathrm{diag}\rbrackets{+2/3, -1/3, +2/3, 0, 0, -1/3}$ related to the electromagnetic charge matrix  taking into account
our choice of physical fields.  This yields the homogeneous Ward identity 
\begin{equation}
 \langle \partial^b_\mu J_\mu^C(x) J_\nu^C(y) \rangle - a^{-3} \partial^b_\mu \delta_{\mu \nu} \delta_{xy} \langle S_{\nu}(y) \rangle = 0
\label{eq:WI}
\end{equation}
at non-zero lattice spacing with the backward lattice derivative $\partial^b_\mu$ and the point-split vector current 
\begin{equation}
 J^C_\mu(x) = \frac{1}{2}\,\sbrackets{ \chibar(x)\,(\gamma_\mu - r)\,U_\mu(x)\,Q_{\rm em}\,\chi(x+a\hmu) +   \chibar(x+a\hmu)\,(\gamma_\mu +
r)\,U_\mu(x)^\dagger\,Q_{\rm em}\chi(x) }\,,
 \label{eq:jc} 
\end{equation}
where  the multiplet $\chi$ collects all flavor components of the three doublets.  The field $S_\nu$ in the contact term in Eq.~(\ref{eq:WI}) reads
 \begin{equation}   
S_\nu(y) = \frac{1}{2}\,\sbrackets{   \chibar(y)\,(\gamma_\nu - r)\,U_\nu(y)\,{Q_{\rm em}}^2\,\chi(y+a\hnu)
- \chibar(y+a\hnu)\,(\gamma_\nu + r)\,U_\nu(y)^\dagger\,{Q_{\rm em}}^2\,\chi(y) } \,. 
\label{eq:def_jc_s}
 \end{equation} 
Thus, the transverse polarization tensor is given by
\begin{align}   \
\Pimn^C(x,y) &= \brackets{J^C_{\mu}(x)\,J^C_{\nu}(y)} - a^{-3}\,\deltamunu\,\delta_{xy}\,\brackets{S_\nu(y)}\,.  
\label{eq:def_pimnc}
\end{align}

 Here, we will first investigate the local variant of the vector current and its  correlation functions. Its interpolating field is given by the usual
quark bilinear,
 \begin{align} 
J^L_\mu(x) &= \chibar(x)\,\gamma_\mu \,Q_{\rm em}\,\chi(x)\,.
 \label{eq:def_jl}  \end{align}
 and we define the bare polarization tensor in position space by the 2-point current correlator 
\begin{align} \Pimn^L(x,y) &= \brackets{J^L_{\mu}(x)\,J^L_{\nu}(y)} \,. 
 \label{eq:def_pimnl}
 \end{align} 
In contrast to the conserved point-split vector current in Eq.~(\ref{eq:jc}), the local vector current is not exactly conserved at
non-zero lattice spacing and hence the polarization tensor $\Pimn^L$ is not transverse.
Therefore, the latter will have to be  potentially additively
and multiplicatively renormalized.
This will be partly discussed later on.

 The polarization tensor $\Pi_{\mu \nu}(Q)$ in momentum space
at Euclidean momentum $Q$ is obtained via the Fourier transform 
\begin{align}   
\Pi_{\mu \nu}(Q) &= a^4\,\sum\limits_{x}\,\epow{iQ\cdot(x+a\hmu/2-y-a\hnu/2)}\,\Pimn(x,y) 
\label{eq:ft_pimn} 
\end{align} 
with spacetime arguments in the Fourier phase shifted by half a lattice spacing.  The polarization function $\Pi(Q)$ is derived from 
$\Pimn(Q)$ using the
projector $P_\munu(Q)$  on the transverse part of the tensor, 
\begin{align}   
P_\munu(Q) &= \Qhat_\mu\,\Qhat_\nu - \deltamunu\,\Qhat^2 \nonumber\\ 
\Pi(Q) &=  \frac{1}{3\,(\Qhat^2)^2} P_\munu(Q) \,\Pimn(Q) \; .   
\label{eq:def_pi}
 \end{align}
$\Qhat$ are the lattice momenta, component-wise related to $Q$ via $\Qhat_\mu = 2 \sin(a Q_\mu / 2) / a$.

  Starting from Eq.~\refeq{eq:def_pi} we define the real and momentum-averaged polarization function
 \begin{align}  
\Pi^{(\mathrm{av})}(\Qhat^2) &= \Real{ \frac{1}{\#\mathcal{G}(Q)}\,\sum\limits_{Q'\,\in\,\mathcal{G}(Q)}\,\Pi(Q') }\,. 
\label{eq:def_pi_avg} 
\end{align}
By explicitly taking the real part, we project on isospin symmetry sectors. This will be further discussed in
Sect.~\ref{sec:symmetry_projection}.
$\mathcal{G}(Q)$ is the set which
contains all momenta obtained from $Q$ applying discrete rotations and  reflections of the 4-dimensional lattice.  We include
rotations mixing time and
spatial coordinates, whenever they are possible, although our configurations  feature $T = 2\,L$ for the lattice time and spatial extent
$T$ and $L$,
respectively.  Moreover, in practice we also average over momenta with the same $\Qhat^2$ which are only  connected  by a spacetime transformation in
the continuum.  Correspondingly, $\#\mathcal{G}(Q)$ denotes the number of elements of this set.  This defines our method to extract the 
scalar vacuum
polarization function as a function of the squared lattice 4-momentum.

  Relations Eq.~\refeq{eq:isospin_1} and Eq.~\refeq{eq:isospin_2} to be given
below show that it is not necessary to calculate the polarization tensor for all combinations of single flavor quark currents as
suggested by
Eqs.~\refeq{eq:jc} and \refeq{eq:def_pimnc}.  It is sufficient to restrict to combinations of single quark currents with, say, plus
components of the quark doublets.

In the following we restrict the discussion to the light valence quark sector.
In the heavy valence sector analogous arguments are used and the latter will be covered in a more general framework in \cite{mp:xxxx}.

\section{Procedure}
\label{sec:procedure}
Many applications of the vacuum polarization function require the vacuum polarization function
in momentum space $\Pi(\Qhat^2)$.
However, below we will investigate
the scaling properties of the polarization tensor
implied by relation~\refeq{eq:def_pi_avg} in position space. This will enable us to draw conclusions on $\Pi(\Qhat^2)$.

Given the on-shell $\order{a}$ improvement of the vector current correlator at physical distances in the continuum limit
~\cite{Frezzotti:2003ni} we focus on the impact of contributions to the Fourier sum from small and zero distance.
Formally, we are interested in the quantity
\begin{align}
  \Pi^{(\mathrm{av})} (\Qhat^2) &= \sbrackets{\frac{1}{3\,(\Qhat^2)^2} P_\munu(Q)\,
    a^4\,\sum\limits_{x\,\in\,V}\,\epow{iQ\cdot(x+a\hmu/2-y-a\hnu/2)}\,\Pimn(x,y)
  }^{\mathrm{av}}
\end{align}
for a physical 4-volume $V$.
This $\Pi^{(\mathrm{av})}$ can be expanded in the continuum limit as
\begin{align}
  \Pi^{(\mathrm{av})} &= \sum\limits_{k\ge -6}\,C_k\,a^k\,,
\end{align}
such that 
\begin{align}
  C_0 = & \frac{1}{3(Q^2)^2} P_\munu(Q)\,\int\limits_{V}\,d^4x\,\brackets{J_\mu(x)\,J_\nu(y)}\,\epow{iQ(x-y)}  \nonumber \\
  & + \mathrm{~contributions~from~operators~of~dimension~6}\,.
\end{align}
We will argue that $C_{1} = 0$ automatically in tmLQCD at maximal twist, irrespective of the remaining $C_k$ for $k 
\ne 1$.

To that end we proceed in two steps:
\begin{enumerate}
  \item We examine the possible mixing of the polarization tensor in position space with operators of equal and lower dimension due to
renormalization and short distance contributions.
    The occurrence of such a mixing requires the definition of a subtracted operator.
  \item We use the Symanzik expansion technique with reference to the twisted mass lattice action and the subtracted operator to
    show that all contributions to $C_{1}$ vanish at maximal twist.
\end{enumerate}

At the non-perturbative level the identification of the mixing pattern and of the terms in the Symanzik expansion relies on the symmetries
of the lattice and the continuum theory. Automatic $\order{a}$ improvement means that no improvement coefficients are 
needed in tmLQCD in order to eliminate $\order{a}$ terms.
The only parameter ultimately assumed to be
tuned is the twist angle such that maximal twist is realized. 
See Refs.~\cite{Frezzotti:2005gi,Baron:2010bv, Baron:2010th} for details how this has been  
achieved for the $N_f=2+1+1$ setup we are interested in here. For our purposes, we only need to recall
that maximal twist corresponds to having a vanishing bare quark mass $m_q=0$ in the Wilson Dirac operator such that the twisted mass
$\mu_q$ takes the
role
of the physical one.

\section{Symmetry projections}
\label{sec:symmetry_projection}

Our discussion of operator mixing and the Symanzik expansion given below proceeds in position space, yet 
the position space current correlators given in Eq.~\refeq{eq:def_pimnc} and \refeq{eq:def_pimnl} do not have a
definite transformation behavior under the symmetries of the lattice theory. To remedy this shortcoming, our definition
of the hadronic vacuum polarization function in
momentum space given in \Eq{eq:def_pi_avg}
incorporates projections on the spacetime symmetry sector
as well as on the isospin symmetry sectors by taking the real part. Since for the following discussion it is desirable
to have definite
transformation
properties in position space as well, we show in this section that the projections defined in momentum space
automatically
imply the corresponding properties for the correlators in position space.

\paragraph{Spacetime transformation group}
The momentum projector $P_\munu(Q)$ given in \Eq{eq:def_pi} transforms like a rank-2  tensor. Restricting the 
set of momenta to a representative set we can extend the average over $\mathcal{G}(Q)$ 
to the complete spacetime transformation group. As outlined in appendix~\ref{app:spacetime_proj} we can
realize this average equivalently in position space. This amounts to defining the projected polarization tensor
\begin{align}
\sbrackets{\Pi_{\mu'\nu'}(x', y')}^{(av)} = \frac{1}{N_\mathcal{G}}\,
  \sum\limits_{R \in \mathcal{G}}\,\Lambda(R)_\mu^{\mu'}\,\Lambda(R)_\nu^{\nu'}\,
  \Pi_\munu(\Lambda(R) x',\Lambda(R) y')\,,
\label{eq:def_Piav_pos}
\end{align}
where $\Lambda(R)$ are the representation matrices of the lattice rotations and reflections.
In this form the vacuum polarization tensor in position space exhibits the transformation
behavior of a rank-2 tensor. We will leave out
the brackets $\sbrackets{\quad}^\mathrm{(av)}$ from position space operators and assume
this exact rank-n tensor transformation behavior for all operators in the following sections.

In anticipation of the following discussion we note, that in particular we have invariance of
the tensor under spacetime inversion $Q \to -Q$ or $x \to -x$. This is one of the key transformations
in the discussion of automatic $\order{a}$ improvement. Moreover, with the definition in Eq.~\refeq{eq:def_Piav_pos}
the average over momentum orbits becomes trivial as in the continuum.

\paragraph{Isospin}
For $SU(2)$ isospin relations
we use the flavor matrices $\tau^{\pm},\,\tau^3$ based on the Pauli matrices,
and $\tau^0 = \mathbb{1}$. Correspondingly, with $J^\tau = \chibar\,\gamma_\mu \tau\,\chi$ we denote the isospin component of the 
current for any of the three doublets.

The implications of taking the real part of the polarization tensor in momentum space can be immediately seen by using the 
relation
\begin{align}
  \brackets{J_\mu^{f_1}(x)\,J_\nu^{f_2}(y)}^* &= \brackets{J_\mu^{\fbar_2}(x)\,J_\nu^{\fbar_1}(y)}
  \label{eq:isospin_1}
\end{align}
of the current correlator in position space and the corresponding relation
\begin{align}
  \Pi_\munu^{f_1 f_2\,*}(Q) &= \Pi_\munu^{\fbar_2 \fbar_1}(-Q)
  \label{eq:isospin_2}
\end{align}
for the polarization tensor in momentum space. Here  $(f_1, f_2)$ denotes a pair of quark flavor indices
and the index with a bar $\fbar_{1/2}$ denotes the flavor with opposite sign of the twisted mass parameter compared to flavor
$f_{1/2}$.

Given the electromagnetic charge matrix we can split the electromagnetic current of the light quarks into its irreducible isospin
components
\begin{align}
  J^{\mathrm{em}}_l = \frac{2}{3}\,J^\mathrm{up} - \frac{1}{3}\,J^\mathrm{down} =  \frac{1}{6}\,J^{\tau^0} + \frac{1}{2}\,J^{\tau^3}\,.
  \label{eq:isospin_3}
\end{align}
Hence, we only need the components with flavor structure $\tau^0$ and $\tauthree$. Using the relation~\refeq{eq:isospin_1} the correlator of two such
isospin currents $J^{a,b} = J^{f} + \sigma_{a,b}\,J^{\fbar}$
with $\sigma_{a,b} \,\in\, \{\pm 1\}$ in momentum space can be decomposed according to
\begin{align}
  \Pi^{ab}_\munu (Q) & = \langle J^{a} J^{b} \rangle \nonumber \\
  &= \Pi_\munu^{f f}(Q) + \sigma_b\,\Pi_\munu^{f\fbar}(Q) + \sigma_a\,\Pi_\munu^{\fbar f}(Q)
    + \sigma_a\sigma_b\,\Pi_\munu^{\fbar \fbar}(Q)\nonumber\\
  &= \Pi_\munu^{f f}(Q) + \sigma_b\,\Pi_\munu^{f \fbar}(Q) + \sigma_a\,\Pi_\munu^{f \fbar\,*}(-Q) 
    + \sigma_a\sigma_b\,\Pi_\munu^{f f\,*}(-Q)\nonumber\\
    &\xrightarrow[]{\sbrackets{\quad}^\mathrm{(av)}}
    2\,\Real{\sbrackets{\Pi_\munu^{f f}(Q)}^\mathrm{(av)}}\,\rbrackets{1 + \sigma_a\sigma_b}
    + 2\,\Real{\sbrackets{\Pi_\munu^{f \fbar}(Q)}^\mathrm{(av)} }\,\rbrackets{\sigma_a + \sigma_b}\nonumber\\
   &\qquad+ 2i\,\Imag{\sbrackets{\Pi_\munu^{f f}(Q)}^\mathrm{(av)}}\,\rbrackets{1 - \sigma_a\sigma_b}
    + 2i\,\Imag{\sbrackets{\Pi_\munu^{f \fbar}(Q)}^\mathrm{(av)} }\,\rbrackets{-\sigma_a + \sigma_b}\,.
  \label{eq:11}
\end{align}
As before, $\sbrackets{\quad}^\mathrm{(av)}$ denotes the average over equivalent momenta, in particular averaging over $Q$ and $-Q$.
From Eq.~\refeq{eq:11} we find that the contributions from the current-current correlator with equal isospin components
for both currents are purely real ($\sigma_a = \sigma_b$), whereas the mixed isospin combinations are
purely imaginary ($\sigma_a = -\sigma_b$). The latter contributions are isospin symmetry breaking lattice artefacts in tmLQCD
as can be checked by symmetry arguments along the lines of the following sections. Retaining only the real part of the averaged momentum 
space correlator removes these terms
explicitly. We thus only need to consider the correlators $\brackets{J^\tau\,J^\tau}$ with $\tau\in \{\mathbb{1}, \tauthree\}$.

Knowing that we only need to consider correlators of same isospin, we can infer, that in position space we always get
correlators for flavor pairs $(f_1,\,f_2)$, which are symmetrized in the indices $(1,2)$ and the bar operation.
These combinations, too, are manifestly real.

Finally, the operator in the contact term \Eq{eq:def_jc_s} contains the squared electromagnetic charge matrix.
Thus, it also consists of two isospin components given by $\tau^0$ and $\tauthree$.
Again the isospin component $\tauthree$ is purely imaginary whereas the component with $\tau^0$ is purely real.
Thus, for the contact term we  may limit our considerations to the component with $\tau^0 = \mathds{1}$.

\section{Mixing of the polarization tensor}
\label{sec:mixing_of_polarization_tensor}

We start our considerations with the local vector current correlator given in Eq.~(\ref{eq:def_pimnl}),
which is symmetry projected as described in the previous section and in Eq.~(\ref{eq:def_pi_avg}). When renormalizing
the vacuum polarization it will in general
mix with operators of equal and lower dimension possessing
the same symmetry transformation properties.
Moreover, Fourier sums on the lattice and the Fourier integrals in the Symanzik effective theory extend over all distances of operator
products. This can give rise to additional terms that need to be subtracted. They are accounted for
by allowing additional contributions
of contact terms, again of equal and lower dimension and with same transformation properties.

The polarization tensor in position space is of mass dimension 6.
We thus write a subtracted polarization tensor in position space as
\begin{align}
  \sbrackets{J^\tau_\mu(x)\,J^\tau_\nu(y)}_\mathrm{sub} 
  &= \sum\limits_{k=0}^6\,\sum\limits_{i\ge 0}\,\frac{Z^{(0)}_{ki}}{a^{6-k}}\,O_{ki\,\munu}(x,y)
    + a^{-4}\,\delta_{x y}\,\sum\limits_{k=0}^6\,\sum\limits_{i\ge 0}\,\frac{Z^{(1)}_{ki} }{a^{2-k}}\,B^{(1)}_{ki\,\munu}(y)\nonumber\\
    &\quad + a^{-4}\,\partialbar^{(x)}_\mu\,\delta_{x y}\,\sum\limits_{k=0}^6\,\sum\limits_{i\ge 0}\,\frac{Z^{(2)}_{ki}}{a^{1-k}}\,B^{(2)}_{ki\,\nu}(y)\nonumber\\
    &\quad + a^{-4}\,\partialbar^{(x)}_\kappa\,\partialbar^{(x)}_\lambda\,\delta_{x y}\,\sum\limits_{k=0}^6\,\sum\limits_{i\ge
0}\,\frac{Z^{(3)}_{ki}}{a^{-k}}\,B^{(3)}_{ki\,\mu\nu\kappa\lambda}(y) \nonumber\\
  &\quad + \ldots \,.
  \label{eq:mixing_1}
\end{align}
With index $k$ we label the dimension of the operators and index $i$ runs over the possible operators within each dimension.
As a lattice version of the Dirac $\delta$ function we use
$a^{-4}\,\delta_{x y} \xrightarrow[] {a \to 0} \delta(x-y)$.
The parity-odd first lattice derivative $\partialbar_\mu$ is given by $\left(\partial^f_\mu + \partial^b_\mu\right)/2$ with $\partial^f_\mu$ and
$\partial^b_\mu$ being the lattice forward and backward partial derivatives, respectively.
For definiteness we  have set $O_{ki\,\munu} = J^\tau_\mu\,J^\tau_\nu$ for $k = 6,\,i = 0$.

When enumerating the operators $O_{ki},\,B^{(n)}_{ki}$, we keep explicit
factors of Wilson and twisted quark mass, $m_q$ and $\mu_q$, respectively, as well as of the dimensionless Wilson parameter $r$
at zeroth and first power. With the parametrization in Eq.~\refeq{eq:mixing_1}, i.e. the explicit factoring out
of powers of the lattice spacing and of quark masses, the dimensionless coefficients
$Z^{(n)}_{ki}$ do not have a power dependence on the lattice spacing~\cite{Luscher:1996sc, Weisz:2010nr}.
The detailed form of these factors would be fixed by a proper set of renormalization conditions. 
We will not formulate such conditions, but stay on the level of a general subtracted operator. This is sufficient
for our purposes, since we are primarily interested in the transformation properties of the operators.

Taking the Fourier transform of \Eq{eq:mixing_1}, the contributions from the operators $B^{(1)}$ are momentum independent,
while those from $B^{(2)}$ and $B^{(3)}$ generate terms that depend on the external momentum.
For $B^{(2)}$ there are no operators to give rise to $\order{a}$ terms.
The general notation for $B^{(3)}_{ki\,\mu\nu\kappa\lambda}$ is meant to include various Lorentz structures,
$B^{(3)}_{ki\,\mu\nu\kappa\lambda} \propto 
B^{(3)}_{ki}\,\delta_{\mu\nu}\,\delta_{\kappa\lambda},\,
B^{(3)}_{ki}\,\delta_{\mu\kappa}\,\delta_{\nu\lambda},\,
B^{(3)}_{ki\,\mu\kappa}\,\delta_{\nu\lambda},\,\mathrm{~etc.}$.
The sets of operators for the $B^{(n)}$
that can mix with the polarization tensor via short-distance
contributions can be constructed from the mass parameters, the Wilson parameter $r$, quark bilinears
and products of those as well as the lattice covariant derivative
and the lattice gauge field strength tensor $C_{\mu \nu}$ for which the expression given in~\cite{Sheikholeslami:1985ij}
can be taken.
The set is restricted by the symmetries of the lattice theory.
For twisted mass lattice QCD we use the following list of symmetry transformations,
\begin{itemize}
  \item twisted time reversal
  \item twisted parity
  \item charge conjugation
  \item $\mathcal{P} \times \mathcal{D}\times \sbrackets{m_0 \to -m_0} \times \sbrackets{r \to -r}$
  \item $\ronetwofive \times \mathcal{D} \times \sbrackets{\mu_q \to -\mu_q}$
\end{itemize}
The details of these transformations are described in~\cite{Shindler:2007vp,Constantinou:2013zqa} and
for completeness a brief listing is given in appendix \ref{app:2}.

To investigate the mixing pattern for $\Pi_\munu^L(x,y)$ obtained from the correlator of two local vector currents in the continuum limit, we
distinguish the two cases
$x = y$ and $x \ne y$ for the spacetime arguments in the Fourier sum
 \begin{align}  
\Pimn^L(Q) &= a^4\,\sum\limits_{x \ne y}\,\brackets{ \sbrackets{J^L_\mu(x)}_R\,\sbrackets{J^L_\nu(y)}_R}\,\epow{iQ(x-y)}     + 
a^4\,\brackets{
\sbrackets{J^L_\mu(y)\,J^L_\nu(y)}_R}\nonumber \\ 
  &= \Pimn^{(2)}(Q) + \Pimn^{(4)}\,,
  \label{eq:mixing_2}
 \end{align}
where $\sbrackets{\quad}_R$ denotes a given renormalization scheme.
 The two terms in \Eq{eq:mixing_2}
have to be considered individually due to their different behavior under renormalization in the continuum  limit.

\setcounter{paragraph}{0}
\paragraph{$x = y$}
$\Pimn^{(4)}$ is the lattice vacuum expectation value of a four-quark operator of mass
dimension 6. We recall, that $\tau$ is either $\tau^0$ or $\tauthree$. 
Additional $\order{a}$ terms and terms with negative powers of the lattice spacing
can also arise through singularities in the limit $x \to y$
when performing the continuum limit in the Symanzik effective theory.
In the continuum these terms can be identified by expanding the operator product to have the form of a ratio $\brackets{O^{(k)}(y)} / ||x-y||^k$
of a condensate over a power of the distance $||x-y||^k$ with $k$ a positive integer (up to logarithms).
These contributions emerge when applying the Fourier
transform over a region extending to one lattice spacing around $y$.

We capture these short-distance contributions by subtracting from the current-current correlator in position space all possible
local operators of equal and lower dimension, which are allowed to appear constrained by the lattice symmetries.
This involves contributions in the form of the $B^{(n)}_{ki}$ given in Eq.~\refeq{eq:mixing_1}. 
The candidate mixing operators $B^{(n)}_{ki}$ have been separated into those that include and do not include
covariant derivatives. They are listed in tables~\reftab{tab:3a}, \reftab{tab:3b}, \reftab{tab:4a} and~\reftab{tab:4b}
in appendix~\ref{app:3}.

\paragraph{$x \ne y$}
$\Pimn^{(2)}(Q)$ is composed of a product of two vector currents in position space at non-zero distance $x \ne y$. This makes
the situation rather definite here. For this operator there is neither mixing nor additive renormalization.
The local current operators are normalized multiplicatively with a factor $Z_V$, which can be determined non-perturbatively~\cite{Martinelli:1994ty}
in a lattice calculation. Thus,
 \begin{align}   
\sbrackets{J^L_\mu(x)}_R &= Z_V\,J^L_\mu(x) \\
   \sbrackets{\Pimn^{(2)}(Q)}_R & = a^4\sum\limits_{x \ne y}\,\sbrackets{J^L_\mu(x)}_R \sbrackets{J^L_\nu(y)}_R\,\,\epow{iQ(x-y)}\,.
\label{eq:mixing_3}
 \end{align}
In the language of Eq.~\refeq{eq:mixing_1} we have $Z_{ki} \ne 0$ only for $(k = 6,\,i=0)$ and zero else.
For automatic $\order{a}$ improvement of the latter correlator for physical distances $x \ne y$ the on-shell
improvement conditions are sufficient within tmLQCD at maximal twist~\cite{Frezzotti:2003ni}.

\section{Symanzik expansion for the local case}
\label{sec:Symanzik_expansion_local_case}
The operators allowed in the mixing pattern when using the local light quark current $J^{
L}_{\mu}(x)$ are listed in tables
\reftab{tab:3a}, \reftab{tab:3b}, \reftab{tab:4a} and~\reftab{tab:4b} in appendix \ref{app:3}.  According to this 
collection the
subtracted operator reads 
\begin{align}  
&\sbrackets{J^\tau_\mu(x)\, J^\tau_\nu(y)}_{\mathrm{sub}}
  = J^\tau_\mu(x)\,J^\tau_\nu(y)
    + \frac{Z^\mathbb{1}}{a^6}\,\deltamunu\,\delta_{xy}
    + \frac{Z^{rm}\,rm_q}{a^5}\,\deltamunu\,\delta_{xy}
    + \frac{Z^{m^2}\,m_q^2 + Z^{\mu^2}\,\mu_q^2}{a^4}\,\deltamunu\,\delta_{xy}\nonumber\\
  &\qquad + \frac{Z^{r\chibar\chi}}{a^3}\,r\,\chibar\chi\,\deltamunu\,\delta_{xy}
    + \frac{Z^{rm^3}\,rm_q^3}{a^3}\,\deltamunu\,\delta_{xy}\nonumber\\
  &\qquad + \frac{1}{a^4}\rbrackets{ Z^{Q^2}\,\deltamunu\,\partialbar^2 + Z^{QQ}\,\partialbar_\mu\,\partialbar_\nu}\,\delta_{xy}
    + \frac{r m_q}{a^3}\rbrackets{ Z^{rmQ^2}\,\deltamunu\,\partialbar^2 + Z^{rmQQ}\,\partialbar_\mu\,\partialbar_\nu}\,\delta_{xy} \nonumber\\
  &\qquad + \mathrm{~operators~of~dimension~} \ge 4   \; .
\label{eq:14.22r}
 \end{align}
 The expansion of the lattice action close to the continuum limit follows from the local effective action
\begin{equation}  
 S_{eff} = S_4 + a S_5 + a^2 S_6 + a^3 S_7 + \ldots \, ,  
 \label{eq:effaction}
\end{equation}
 where $S_k \equiv \int \Llagrange_k \ d^4x$ and the terms
$\Llagrange_k$ contain linear combinations of fields with mass dimension $k$. We expand its exponential up to $\mathcal{O}(a^3)$.
The corrections to  the gauge field Lagrangian in the continuum limit
start with $\order{a^2}$ and in fact contain only even powers of the  lattice spacing~\cite{Luscher:1984xn}. We thus concentrate on the corrections to
the fermion action.  The operators that can appear in $\Llagrange_5$ and $\Llagrange_6$ have been listed in
Refs.~\cite{Sheikholeslami:1985ij, Luscher:1996sc}.  

From the expansion of the operator \Eq{eq:14.22r} and $\exp{(-S_{eff})}$ in \Eq{eq:effaction} the full Symanzik expansion in momentum space is
obtained and reads 
\begin{align}   
%%% a^4\,\brackets{\sbrackets{W^\tau_\munu}_{\mathrm{sub}}}
 \Pi^\tau_\munu(Q) &=
a^4\,\brackets{J^\tau_\mu(y)\,J^\tau_\nu(y)}_0
%%% order a^-2  
 + \frac{\tilde{Z}^\mathbb{1}}{a^2}\,\deltamunu\nonumber\\   
%%% order a^-1   
&\quad   + \frac{\tilde{Z}^\mathbb{1}}{a}\,\brackets{-S_5}_0\,\deltamunu   + \frac{\tilde{Z}^{rm}\,rm_q}{a}\,\deltamunu\nonumber\\
   %%% order a^0  
&\quad   + \tilde{Z}^\mathbb{1}\,\brackets{-S_6 + \frac{1}{2} S_5^2}_0\,\deltamunu   + \tilde{Z}^{rm}\,\brackets{-r m_q\,S_5}_0\,\deltamunu   +
\rbrackets{\tilde{Z}^{m^2}\,m_q^2 + \tilde{Z}^{\mu^2}\,\mu_q^2}\,\deltamunu\nonumber\\
%%% order a   
&\quad   + a\,\tilde{Z}^\mathbb{1}\,\brackets{-S_7 + S_5 S_6 - \frac{1}{6} S_5^3}_0\,\deltamunu   +
a\,\brackets{\rbrackets{\tilde{Z}^{rm}\,rm_q}\, \left (- S_6 + \frac{1}{2} S_5^2 \right )}_0\,\deltamunu \nonumber\\
&\quad + a\,\brackets{-\rbrackets{\tilde{Z}^{m^2}\,m_q^2 +
\tilde{Z}^{\mu^2}\,\mu_q^2}\,S_5}_0\,\deltamunu    + a\,\tilde{Z}^{r\chibar\chi}\,\brackets{r\,\chibar\chi}_0\,\deltamunu 
+ a\,\tilde{Z}^{rm^3}\,r\,m_q^3\,\deltamunu\nonumber\\
%%% momentum dependend    
&\quad + \rbrackets{ \tilde{Z}^{Q^2}\,\deltamunu\,\Qhat^2 + \tilde{Z}^{QQ}\,\Qhat_\mu\,\Qhat_\nu}
+ a r m_q\,\rbrackets{ \tilde{Z}^{rmQ^2}\,\deltamunu\,\Qhat^2 + \tilde{Z}^{rmQQ}\,\Qhat_\mu\,\Qhat_\nu} \nonumber\\
%%% 
&\quad + \left\{ \order{a^2}\,, \mathrm{~operators~of~higher~dimension~} \right\}\,.
  \label{eq:14.22t}
\end{align}
Since we are working at maximal twist $m_q \to 0$, we may drop all terms involving the untwisted quark mass.
  Using the $\ronetwofive$-symmetry~\cite{Shindler:2007vp} we see that the vacuum
expectation values $\brackets{\; }_0$ of  $S_5$, $S_5 S_6$ as well as of $\mu^2_qS_5$ and $\chibar\chi$ vanish as these merely
contain 
$\ronetwofive$-odd  operators. 
Similarly all terms in $S_7$ disappear by either the $\ronetwofive$- or the $\mathcal{P} \times \sbrackets{\mu_q \to -\mu_q}$ symmetry as 
is demonstrated
in appendix~\ref{app:sseven}.

We may then conclude that at maximal twist there are no $\mathcal{O}(a)$ lattice artefacts stemming
from the contributions in Eq.~\refeq{eq:14.22t} to $\Pi^\tau_\munu$,
whose Symanzik expansion we write again for this case,
\begin{align}
%%% a^4\,\brackets{\sbrackets{W^\tau_\munu}_{\mathrm{sub}}} 
 \Pi^\tau_\munu(Q) &=
  a^4\,\brackets{J^\tau_\mu(y)\,J^\tau_\nu(y)}_0
%%% order a^-2  
  + \frac{\tilde{Z}^\mathbb{1}}{a^2}\,\deltamunu   
%%% order a^0  
  + \tilde{Z}^\mathbb{1}\,\brackets{-S_6 + \frac{1}{2} S_5^2}_0\,\deltamunu
  + \rbrackets{ \tilde{Z}^{\mu^2}\,\mu_q^2}\,\deltamunu \nonumber\\
  &\quad + \rbrackets{ \tilde{Z}^{Q^2}\,\deltamunu\,Q^2 + \tilde{Z}^{QQ}\,Q_\mu\,Q_\nu}
  + \left\{ \order{a^2}\,, \mathrm{~operators~of~higher~dimension~} \right\}\,.
 \label{eq:wmunutm} 
\end{align}

\section{Application to the conserved current correlator}
\label{sec:application_to_conserved_current_correlator}
The conserved current can be written in the following form 
\begin{align}
  J_\mu^C(x) &= J^L_\mu(x) + \frac{a}{2}\,\sbrackets{
\chibar\,\gamma_\mu\,\tau\,     \rbrackets{\overrightarrow{\nabla}_\mu^f + \overleftarrow{\nabla}_\mu^f}\,\chi}(x)    - \frac{ar}{2}\,\sbrackets{
\chibar\,\tau\,     \rbrackets{\overrightarrow{\nabla}_\mu^f - \overleftarrow{\nabla}_\mu^f}\,\chi}(x)\,,  
\label{eq:rel_jc_jl} 
\end{align}
where $\overrightarrow{\nabla}_\mu^f$ is the covariant forward lattice derivative acting to the right.
 It is a sum of the local current operator and two local operators of mass dimension 4. 
  Similarly, for the field in the lattice contact term in Eq.~(\ref{eq:def_pimnc}) we have 
\begin{align}   
   S^\tau_\nu(y) &= \frac{a}{2}\,\sbrackets{ \chibar\,\tau\,\gamma_\nu\,    
\rbrackets{\overrightarrow{\nabla}_\nu^f - \overleftarrow{\nabla}_\nu^f}\,\chi}(y)   -\frac{ar}{2}\,\sbrackets{ \chibar\,\tau\,    
\rbrackets{\overrightarrow{\nabla}_\nu^f + \overleftarrow{\nabla}_\nu^f}\,\chi}(y)   -r\,\chibar\,\tau\,\chi(y) \, .    
\label{eq:SymExpContact}
 \end{align} 
Hence, both the conserved current as well as the lattice contact term are a sum of local 
quark-bilinear operators for whose correlators we can use 
the 
Symanzik expansion. 

Having written the conserved current as the local current plus two operators containing derivatives that are of  dimension 4 implies that 
there is no principle alteration of the mixing with lower dimensional operators for $\brackets{J_\mu^C\,J_\nu^C}$ compared to 
the local case, since $J_\mu^C\,J_\nu^C$ can be expressed as a sum of the local-current correlator and additional terms of dimension
7  and 
8. Moreover, for
the short-distance part of the vacuum polarization tensor formed from the conserved current
the appearance of mixing operators is further constrained by the vector Ward identity Eq.~(\ref{eq:WI}).
Thus, the considerations for the occurrence of $\order{a}$ terms are basically the same as for the local case. 

The only addition  is the  lattice contact term
where we have $r\chibar\,\tau\,\chi$. As stated earlier, due to the symmetry projections discussed in 
Sect.~\ref{sec:symmetry_projection},  $\chibar\,\tau\,\chi$ with
$\tau=\tauthree$ is excluded and only  $\tau = \mathbb{1}$ needs to be considered. 
At maximal twist, when $\ronetwofive$ is a symmetry of the continuum
theory, this term will vanish, since  it is odd under $\ronetwofive$.

Combining the above arguments, the hadronic vacuum polarization function formed from the conserved vector current according to 
\Eq{eq:def_pimnc}, \Eq{eq:def_pi} and Eq.~(\ref{eq:def_pi_avg}) exhibits no $\mathcal{O}(a)$ contributions.

\section{Conclusions}
\label{sec:conclusions}

A crucial element in obtaining accurate results from lattice QCD calculations
is the suppression of lattice spacing artefacts 
and a controlled approach towards the continuum limit. 
The lattice community has therefore developed a number of actions and 
improved operators that guarantee that physical quantities scale with a rate 
of $\order{a^2}$ to the continuum limit. 

One particular lattice QCD formulation, which we have investigated 
here, is the twisted mass formulation~\cite{Frezzotti:2000nk, Frezzotti:2003ni, Frezzotti:2003xj, Frezzotti:2004wz}. 
When tuning the twisted mass lattice action to maximal twist 
physical quantities are automatically 
$\order{a}$ improved~\cite{Frezzotti:2003ni}. 
Indeed, in numerical computations with two dynamical quarks the $\order{a^2}$ scaling of many 
physical quantities could be demonstrated~\cite{Baron:2009wt,Dimopoulos:2007qy,Alexandrou:2008tn} showing also
that these remaining $\order{a^2}$ lattice artefacts are often 
very small as can be deduced from~\cite{Frezzotti:2005gi}.

However, the arguments that lead to $\order{a}$ improvement
for twisted mass fermions 
do not immediately cover quantities that involve summations over all lattice 
points thus possessing short-distance contributions.

Here, we have examined the behavior of the hadronic vacuum polarization function which 
serves as a most important basic quantity to compute hadronic contributions 
to electroweak observables, quark masses and also the strong coupling constant. 
In order to see whether short-distance contributions affect the 
rate of the continuum limit scaling, we 
have constructed the Symanzik expansion for these short-distance contributions.

We have found that when the theory is tuned to maximal twist,
automatic O(a) improvement prevails for the complete vacuum polarization function provided
that it is defined as eigenstate of the symmetry transformations of the lattice action.
Thus, 
continuum limit extrapolations of our lattice results can safely be performed 
employing fit functions without linear terms in the lattice 
spacing as has been done in~\cite{Burger:2013jya}.
In the course of this work, we have
established the classification of the twisted mass symmetry properties of 
operators up to dimension 6, see  
appendix~\ref{app:3} for a complete list.

In this paper, we have concentrated on the twisted mass formulation of lattice QCD. 
However, it would be important to extend the analysis to other 
lattice formulations of QCD to ensure that the short-distance 
contributions do not spoil the desired $\order{a}$ improvement of 
the corresponding vacuum polarization function. 

Another extension of the present work, which however goes
substantially beyond the scope of this paper, is a potentially generalized analysis of short-distance
contributions to a larger class of operators in twisted mass lattice QCD, which is currently
under investigation \cite{mp:xxxx}.

\section*{Acknowledgments}
We are grateful to K. Cichy and E. Garcia Ramos for 
very constructive discussions on the $\order{a}$ improvement 
of quantities which involve short-distance contributions. 
Special thanks goes to the referee of~\cite{Burger:2013jya} for raising the question about the $\mathcal{O}(a)$ improvement of the vacuum polarization function and thus motivating this work.
This work has been supported in part by the DFG Corroborative
Research Center SFB/TR9.
G.H.~gratefully acknowledges the support of the German Academic National Foundation (Studienstiftung des deutschen Volkes e.V.) and of the
DFG-funded Graduate School GK 1504.
K.J. was supported in part by the Cyprus Research Promotion
Foundation under contract $\Pi$PO$\Sigma$E$\Lambda$KY$\Sigma$H/EM$\Pi$EIPO$\Sigma$/0311/16.

\appendix

\section{Spacetime symmetry projections in position space}
\label{app:spacetime_proj}
The momentum projector $P_\munu(Q)$ given in \refeq{eq:def_pi} transforms like a rank-2-tensor, that is for any 
discrete spacetime transformation $\Lambda$ we have
\begin{align*}
  P_\munu(\Lambda Q) &= \Lambda_\mu^{\mu'}\,\Lambda_\nu^{\nu'}\,P_{\mu'\nu'}(Q)\,.
\end{align*}
$\Lambda$ denotes a representation of the essentially hypercubic lattice symmetry group.
We can restrict the set of momenta to a representative set and translate the average over $\mathcal{G}(Q)$ to position space.
Moreover, instead of averaging over $\mathcal{G}(Q)$ for a specific momentum $Q$ we can average over the complete spacetime
transformation group $\mathcal{G}$
\footnote{For any momentum $Q$ the number of elements $N_{\mathcal{G}(Q)}$ divides the number of elements in the whole
group $N_{\mathcal{G}}$.}%@@ ???
and define
\begin{align}
  \sbrackets{\Pi(Q)}^\mathrm{(av)} &= \frac{1}{N_{\mathcal{G}(Q)}}\,\sum\limits_{Q \in \mathcal{G}(Q)}\,\frac{1}{3(\hat{Q}^2)^2}\,P_\munu(Q)\,
a^4\,\sum\limits_{x}\,\Pi_\munu(x,y)\,\epow{iQ(x+a\hmu/2-y-a\hnu/2)}\nonumber\\
  &\quad = \frac{1}{N_{\mathcal{G}}}\,\sum\limits_{\Lambda \in \mathcal{G}}\,\frac{1}{3(\hat{Q}^2)^2}\,P_\munu(\Lambda Q_\textrm{fix})\,
  a^4\,\sum\limits_{x}\,\Pi_\munu(x,y)\,\epow{i(\Lambda Q_\textrm{fix})(x+a\hmu/2-y-a\hnu/2)}\nonumber\\
%%%
&\quad = \frac{P_{\mu'\nu'} (Q_\textrm{fix})}{3(\hat{Q}^2)^2}\,                        
a^4\,\sum\limits_{x}\,\frac{1}{N_\mathcal{G}}\,
\sum\limits_{\Lambda \in \mathcal{G}}\,
\Lambda_\mu^{\mu'}\,\Lambda_\nu^{\nu'}\,\Pi_\munu(x,y)\,\epow{iQ_\textrm{fix}\,\Lambda^{-1}(x+a\hmu/2-y-a\hnu/2)}
\label{eq:mom_av_in_pos_space}
\end{align}
where $Q_\textrm{fix}$ is some fixed reference momentum. We can rewrite the transformed spacetime argument in the Fourier phase in
Eq.~\refeq{eq:mom_av_in_pos_space} as
\begin{align}
  \Lambda^{-1}(x+a\hmu/2) &= x' + a\hmu'/2 \nonumber\\ 
   \mu' &= \sigma_\Lambda(\mu)\nonumber\\ 
   x' &= \left\{ \begin{matrix} \Lambda^{-1}x & \mu-\mathrm{direction~not~reflected} \\
                 \Lambda^{-1}(x+a\hmu) & \mu-\mathrm{direction~reflected}
   \end{matrix} \right.\,,
  \label{eq:13}
\end{align}
where $\sigma_\Lambda$ is the permutation generated by $\Lambda$. 
Hence, we obtain
\begin{align}
  \sbrackets{\Pi(Q)}^\mathrm{(av)} &=
  \frac{1}{3(\hat{Q}^2)^2}\,P_{\mu'\nu'} (Q_\textrm{fix})\,
    a^4\,\sum\limits_{x'}\,\frac{1}{N_\mathcal{G}}\,
    \sum\limits_{\Lambda \in \mathcal{G}}\,\Lambda_\mu^{\mu'}\,\Lambda_\nu^{\nu'}\,
    \Pi_\munu(\Lambda x',\Lambda y')\,\epow{iQ_\textrm{fix}\,(x'+a\hmu'/2-y'-a\hnu'/2)} \nonumber\\
    &= \frac{1}{3(\hat{Q}^2)^2}\,P_{\mu'\nu'} (Q_\textrm{fix})\,
    a^4\,\sum\limits_{x'}\,
    \sbrackets{\Pi_{\mu'\nu'}(x', y')}^{(av)}\,\epow{iQ_\textrm{fix}\,(x'+a\hmu'/2-y'-a\hnu'/2)} \; . \nonumber\\
  \label{eq:14}
\end{align}
By construction the operator
\begin{equation}
 \sbrackets{\Pi_{\mu'\nu'}(x', y')}^{(av)} = \frac{1}{N_\mathcal{G}}\,
    \sum\limits_{\Lambda \in \mathcal{G}}\,\Lambda_\mu^{\mu'}\,\Lambda_\nu^{\nu'}\,
    \Pi_\munu(\Lambda x',\Lambda y')
\end{equation}
has the same transformation behavior as the
projector $P_\munu$; it transforms like a true rank-2 tensor in position space and the trace of the tensor,
$\sum\limits_{\mu'}\,\sbrackets{\Pi_{\mu' \mu'}(x', y')}^{(av)}$, is a scalar.

\section{Symmetry transformations}
\label{app:2}

\begin{align}
  \mathcal{T}_{1,2} &: & x &\to Tx = (-x_0,\,\xvec) \nonumber\\
  &\hphantom{:} &\chi(x) &\to i\tau^{1,2}\,\gamma_0\,\gammafive\,\chi(Tx) \nonumber\\
  &\hphantom{:} &\chibar(x) &\to -i\chibar(Tx)\,\tau^{1,2}\,\gammafive\,\gamma_0 \nonumber\\
  &\hphantom{:} &U_0(x) &\to U_0(Tx-a\hat{0})^\dagger\,,\; U_i(x) \to U_i(Tx)\nonumber\\
%%%
  \mathcal{T} \times \sbrackets{\mu_q \to -\mu_q} &:& &\nonumber\\
  & \textrm{with} \; \mathcal{T} : & x &\to Tx = (-x_0,\,\xvec) \nonumber\\
  &\hphantom{:} &\chi(x) &\to i\,\gamma_0\,\gammafive\,\chi(Tx) \nonumber\\
  &\hphantom{:} &\chibar(x) &\to -i\chibar(Tx)\,\gammafive\,\gamma_0 \nonumber\\
  &\hphantom{:} &U_0(x) &\to U_0(Tx-a\hat{0})^\dagger\,,\; U_i(x) \to U_i(Tx)\nonumber\\
%  \midrule
  \hline
%%%
%%%
  \mathcal{P}_{1,2} &: & x &\to Px = (x_0,\,-\xvec)\nonumber\\
  &\hphantom{:} &\chi(x) &\to i\tau^{1,2}\,\gamma_0\,\chi(Px)\nonumber\\
  &\hphantom{:} &\chibar(x) &\to -i\chibar(Px)\,\tau^{1,2}\,\gamma_0\nonumber\\
  &\hphantom{:} &U_0(x) &\to U_0(Px)\,,\; U_i(x) \to U_i(Px - a\hat{i})^\dagger\nonumber\\
%%%
  \mathcal{P} \times \sbrackets{\mu_q \to -\mu_q} &:& &\nonumber\\
  & \textrm{with} \; \mathcal{P}: & x &\to Px = (x_0,\,-\xvec)\nonumber\\
  &\hphantom{:} &\chi(x) &\to i\,\gamma_0\,\chi(Px)\nonumber\\
  &\hphantom{:} &\chibar(x) &\to -i\chibar(Px)\,\gamma_0\nonumber\\
  &\hphantom{:} &U_0(x) &\to U_0(Px)\,,\; U_i(x) \to U_i(Px - a\hat{i})^\dagger\nonumber\\
%  \midrule
  \hline
%%%
%%%
%%%
  \mathcal{C} &: &\chi(x) &\to C^{-1}\,\chibar(x)^T\nonumber\\
  &\hphantom{:} &\chibar(x) &\to -\chi(x)^T\,C\nonumber\\
  &\hphantom{:} &U_\mu(x) &\to U_\mu(x)^*\nonumber\\
  & \textrm{with} &  C &=  i \gamma_0 \gamma_2\;  \textrm{in}\; \textrm{representation}\; \textrm{of~\cite{Shindler:2007vp}}\nonumber\\
%  \midrule
  \hline
\end{align}
%%%
%%%
%%%
\begin{align}
  \mathcal{P} \times \mathcal{D}\times \sbrackets{m_0 \to -m_0} \times \sbrackets{r \to -r} &: & &\nonumber\\
  & \textrm{with} \;\mathcal{D}: & U_\mu(x) & \to U_\mu(-x-a\hmu)^\dagger\nonumber\\
  &\hphantom{:} & \chi(x) &\to -i\,\chi(-x) \nonumber\\
  &\hphantom{:}& \chibar(x) &\to -i\,\chibar(-x) \nonumber \\
%  \midrule
  \hline
%%%
%%%
%%%
  \ronetwofive \times \mathcal{D} \times \sbrackets{\mu_q \to -\mu_q} &: & &\nonumber \\
  &\textrm{with} \;\ronetwofive : & \chi(x) &\to i\,\gammafive\,\tau^{1,2}\,\chi(x) \nonumber\\
  &\hphantom{:} & \chibar(x) &\to i\,\chibar(x)\,\gammafive\,\tau^{1,2} \nonumber 
\end{align}

% \begin{align}
%   \ronetwofive\,:\, & \chi(x) \to i\,\gammafive\,\tau^{1/2}\,\chi(x) \nonumber\\
%   & \chibar(x) \to i\,\chibar(x)\,\gammafive\,\tau^{1/2}
%   \label{eq:21}\\
%   \mathcal{D}\,:\, & U_\mu(x) \to U_\mu(-x-a\hmu)^\dagger\nonumber\\
%   & \chi(x) \to -i\,\chi(-x) \nonumber\\
%   & \chibar(x) \to -i\,\chibar(-x)
%   \label{eq:22}
% \end{align}

\newpage

\section{Operator listings}
\label{app:3}

The relevant lattice operators which potentially mix with $\Pi_\munu$ at short distances are listed in the following tables \reftab{tab:3a},
\reftab{tab:3b}, \reftab{tab:4a} and \reftab{tab:4b}. The
first pair contains operators not involving derivatives whereas the second accommodates the derivative operators.
We note that for obtaining a complete set of operators
for any operator $O_\munu$ appearing in the tables the diagonal part $\delta_\munu\,O_{\mu\mu}$ (without summation over $\mu$) and the trace
$\delta_\munu\,O_{\lambda\lambda}$ must be included separately. Since these have the same quantum numbers as $O_\munu$ given in the table (with
$I_{\mu\mu} =1$), we do not repeat those quantum numbers. 

Furthermore, to save space the common prefactor $r^k\,m_q^{n_m}\,\mu_q^{n_\mu}$ ($k \in \{ 0, 1\}$, $n_m,\, n_\mu \in \mathds{N}_0$), which is
essential for counting the dimension of the operator, is omitted for all but the first operator. Its quantum numbers can be inferred from the first
line of each table and have to be multiplied with the quantum numbers in the respective column. For the reader's convenience we have added as
supplementary material an expanded list of non-derivative operators, which contain the operators up to dimension six relevant for the
discussion of
$\order{a}$ improvement.

The powers of $\tauthree$ and $\gammafive$ appearing in fermion bilinears such as 
$$r^k\,m_q^{n_m}\,\mu_q^{n_\mu} \chibar (\tauthree)^m (\gammafive)^l
\Gamma \chi$$ with $\Gamma \in \{ \mathbb{1}, \gamma_\mu, \sigma_{\mu\nu}, \gammafive \gamma_\mu, \gammafive \}$
 and four-quark operators
$$r^k\,m_q^{n_m}\,\mu_q^{n_\mu} \chibar (\tauthree)^m (\gammafive)^l \Gamma \chi \chibar (\tauthree)^{m'} (\gammafive)^{l'}
\Gamma \chi $$
 can take the values $m,\, m', \, l, \, l' \in \{ 0, 1\}$.
\clearpage
%\setmargins  {1.0cm}{1.0cm}% % linker & oberer Rand
%             {20cm}{27cm}%   % Textbreite und -hoehe
%             {10pt}{15pt}%   % Kopfzeilenhoehe und -abstand
%             {0pt}{10pt}%    % \footheight (egal) und Fusszeilenabstand
%
%%%%%%%%%%%%%%%%%%%%%%%%%%%%%%%%%%%%%%%%%%%%%%%%%%%%%%%%%%%%%%%%%%%%%%%%%
% non_derivative_symtranstab_partI.tex
%
% Sun Nov 23 19:11:03 EET 2014
%
% PURPOSE
%%%%%%%%%%%%%%%%%%%%%%%%%%%%%%%%%%%%%%%%%%%%%%%%%%%%%%%%%%%%%%%%%%%%%%%%%
\begin{center}
  { \footnotesize
\begin{longtable}{c|rr|rr}
    operator & $\mathcal{P}_{1,2}$ & $\mathcal{P}\,[-\mu]$ & $\mathcal{T}_{1,2}$ & $\mathcal{T}\,[-\mu]$ \\
  \hline
%%%%%%%%%%%%%%%%%%%%%%%%%%%%%%%%%%%%%%%%%%%%%%%%%%%%%%%%%%%%%%%
% r m mu
%%%%%%%%%%%%%%%%%%%%%%%%%%%%%%%%%%%%%%%%%%%%%%%%%%%%%%%%%%%%%%%
  $\deltamunu\,r^k\,m^{n_m}\,\mu^{n_\mu}$                         & $I_\munu$ & $(-)^{n_\mu}\,I_\munu$ & $I_\munu$ & $(-)^{n_\mu}\,I_\munu$ \\
  \hline
%%%%%%%%%%%%%%%%%%%%%%%%%%%%%%%%%%%%%%%%%%%%%%%%%%%%%%%%%%%%%%%%%%%%%%%%%%%%%%%%%%%%%%%%%%%%%%%%%%%%%%%%%%%%%%%%%%%%%%%%%%%%%%
%%%%%%%%%%%%%%%%%%%%%%%%%%%%%%%%%%%%%%%%%%%%%%%%%%%%%%%%%%%%%%%%%%%%%%%%%%%%%%%%%%%%%%%%%%%%%%%%%%%%%%%%%%%%%%%%%%%%%%%%%%%%%%
%%%%%%%%%%%%%%%%%%%%%%%%%%%%%%%%%%%%%%%%%%%%%%%%%%%%%%%%%%%%%%%%%%%%%%%%%%%%%%%%%%%%%%%%%%%%%%%%%%%%%%%%%%%%%%%%%%%%%%%%%%%%%%
%%%
%%% quark-bilinears
%%%
%%%%%%%%%%%%%%%%%%%%%%%%%%%%%%%%%%%%%%%%%%%%%%%%%%%%%%%%%%%%%%%%%%%%%%%%%%%%%%%%%%%%%%%%%%%%%%%%%%%%%%%%%%%%%%%%%%%%%%%%%%%%%%
%%%%%%%%%%%%%%%%%%%%%%%%%%%%%%%%%%%%%%%%%%%%%%%%%%%%%%%%%%%%%%%%%%%%%%%%%%%%%%%%%%%%%%%%%%%%%%%%%%%%%%%%%%%%%%%%%%%%%%%%%%%%%%
%%%%%%%%%%%%%%%%%%%%%%%%%%%%%%%%%%%%%%%%%%%%%%%%%%%%%%%%%%%%%%%%%%%%%%%%%%%%%%%%%%%%%%%%%%%%%%%%%%%%%%%%%%%%%%%%%%%%%%%%%%%%%%

%%%%%%%%%%%%%%%%%%%%%%%%%%%%%%%%%%%%%%%%%%%%%%%%%%%%%%%%%%%%%%%
% 1
%%%%%%%%%%%%%%%%%%%%%%%%%%%%%%%%%%%%%%%%%%%%%%%%%%%%%%%%%%%%%%%
  $\deltamunu\,\chibar\,(\tau^3)^{m}\,\gammafive^{l}\,\chi(x)$ & $(-)^{m+l}\,I_\munu$ & $(-)^{l}\,I_\munu$ & $(-)^{m+l}\,I_\munu$ & $(-)^{l}\,I_\munu$ \\
  \hline
%%%%%%%%%%%%%%%%%%%%%%%%%%%%%%%%%%%%%%%%%%%%%%%%%%%%%%%%%%%%%%%
% C C
%%%%%%%%%%%%%%%%%%%%%%%%%%%%%%%%%%%%%%%%%%%%%%%%%%%%%%%%%%%%%%%
  $\tr{ C_{\mu\lambda}\,C_{\lambda\nu}(x) }        $   & $I_\munu$ & $I_\munu$ & $I_\munu$ & $I_\munu$ \\
  $\tr{ C_{\mu\lambda}\,\tilde{C}_{\lambda\nu}(x)} + [\mu \leftrightarrow \nu] $   & $-I_\munu$ & $-I_\munu$ & $-I_\munu$ & $-I_\munu$ \\
\end{longtable}
}
\end{center}

%%%%%%%%%%%%%%%%%%%%%%%%%%%%%%%%%%%%%%%%%%%%%%%%%%%%%%%%%%%%%%%%%%%%%%%%%
% non_derivative_symtranstab_partII.tex
%
% Sun Nov 23 19:09:29 EET 2014
%
% PURPOSE
%%%%%%%%%%%%%%%%%%%%%%%%%%%%%%%%%%%%%%%%%%%%%%%%%%%%%%%%%%%%%%%%%%%%%%%%%
\begin{center}
  { \footnotesize
  \begin{longtable}{c|r|r|r}
    operator & $\mathcal{C}$ & $\mathcal{P}\,\mathcal{D}\,[-m]\,[-r]$ & $\ronetwofive\,\mathcal{D}\,[-\mu]$ \\
  \hline
%%%%%%%%%%%%%%%%%%%%%%%%%%%%%%%%%%%%%%%%%%%%%%%%%%%%%%%%%%%%%%%
% r m mu
%%%%%%%%%%%%%%%%%%%%%%%%%%%%%%%%%%%%%%%%%%%%%%%%%%%%%%%%%%%%%%%
  $\deltamunu\,r^k\,m^{n_m}\,\mu^{n_\mu}$ & $+1$ & $(-)^{n_m + k}\,I_\munu$ & $(-)^{n_\mu}$ \\
  \hline
%%%%%%%%%%%%%%%%%%%%%%%%%%%%%%%%%%%%%%%%%%%%%%%%%%%%%%%%%%%%%%%%%%%%%%%%%%%%%%%%%%%%%%%%%%%%%%%%%%%%%%%%%%%%%%%%%%%%%%%%%%%%%%
%%%%%%%%%%%%%%%%%%%%%%%%%%%%%%%%%%%%%%%%%%%%%%%%%%%%%%%%%%%%%%%%%%%%%%%%%%%%%%%%%%%%%%%%%%%%%%%%%%%%%%%%%%%%%%%%%%%%%%%%%%%%%%
%%%%%%%%%%%%%%%%%%%%%%%%%%%%%%%%%%%%%%%%%%%%%%%%%%%%%%%%%%%%%%%%%%%%%%%%%%%%%%%%%%%%%%%%%%%%%%%%%%%%%%%%%%%%%%%%%%%%%%%%%%%%%%
%%%
%%% quark-bilinears
%%%
%%%%%%%%%%%%%%%%%%%%%%%%%%%%%%%%%%%%%%%%%%%%%%%%%%%%%%%%%%%%%%%%%%%%%%%%%%%%%%%%%%%%%%%%%%%%%%%%%%%%%%%%%%%%%%%%%%%%%%%%%%%%%%
%%%%%%%%%%%%%%%%%%%%%%%%%%%%%%%%%%%%%%%%%%%%%%%%%%%%%%%%%%%%%%%%%%%%%%%%%%%%%%%%%%%%%%%%%%%%%%%%%%%%%%%%%%%%%%%%%%%%%%%%%%%%%%
%%%%%%%%%%%%%%%%%%%%%%%%%%%%%%%%%%%%%%%%%%%%%%%%%%%%%%%%%%%%%%%%%%%%%%%%%%%%%%%%%%%%%%%%%%%%%%%%%%%%%%%%%%%%%%%%%%%%%%%%%%%%%%

%%%%%%%%%%%%%%%%%%%%%%%%%%%%%%%%%%%%%%%%%%%%%%%%%%%%%%%%%%%%%%%
% 1
%%%%%%%%%%%%%%%%%%%%%%%%%%%%%%%%%%%%%%%%%%%%%%%%%%%%%%%%%%%%%%%
  $\deltamunu\,\chibar\,(\tau^3)^{m}\,\gammafive^{l}\,\chi(x)$ & $+1$ & $(-)^{l+1}\,I_\munu$ & $(-)^{m}$ \\
  \hline
%%%%%%%%%%%%%%%%%%%%%%%%%%%%%%%%%%%%%%%%%%%%%%%%%%%%%%%%%%%%%%%
% C C
%%%%%%%%%%%%%%%%%%%%%%%%%%%%%%%%%%%%%%%%%%%%%%%%%%%%%%%%%%%%%%%
  $\tr{ C_{\mu\lambda}\,C_{\lambda\nu}(x) }        $   & $+1$ & $I_\munu$ & $+1$ \\
  $\tr{ C_{\mu\lambda}\,\tilde{C}_{\lambda\nu}(x)} + [\mu \leftrightarrow \nu] $   & $+1$ & $-I_\munu$ & $+1$ \\
\end{longtable}
}  % of fontsize
\end{center}

%%%%%%%%%%%%%%%%%%%%%%%%%%%%%%%%%%%%%%%%%%%%%%%%%%%%%%%%%%%%%%%%%%%%%%%%%
% derivative_operator_symtranstab_partI.tex
%
% Sun Nov 23 19:01:18 EET 2014
%
% PURPOSE
%%%%%%%%%%%%%%%%%%%%%%%%%%%%%%%%%%%%%%%%%%%%%%%%%%%%%%%%%%%%%%%%%%%%%%%%%
\begin{center}
  {\footnotesize
  \begin{longtable}{c|rr|rr}
     operator  &  $\mathcal{P}_{1,2}$  &  $\mathcal{P}\,[-\mu]$  &  $\mathcal{T}_{1,2}$  &  $\mathcal{T}\,[-\mu]$ \\
\hline
$r^k m^{n_m} \mu^{n_{\mu}}$  &  +1  &  $(-)^{n_\mu}$  &  +1  &  $(-)^{n_\mu}$ \\
\hline
%%%%%%%%%%%%%%%%%%%%%%%%%%%%%%%%%%%%%%%%%%%%%%%%%%%%%%%%%%%%%%%
% r_mu r_kappa C_kappa_nu
%%%%%%%%%%%%%%%%%%%%%%%%%%%%%%%%%%%%%%%%%%%%%%%%%%%%%%%%%%%%%%%
      $ \Tr{ \nablar_\mu\,\nablar_\lambda\,C_{\lambda\nu}(x) + \nablar_\nu\,\nablar_\lambda\,C_{\lambda\mu}(x) }$  &  $I_\munu$  &  $I_\munu$  &  $I_\munu$  &  $I_\munu$ \\
      $ \Tr{ \nablar_\mu\,\nablar_\lambda\,\tilde{C}_{\lambda\nu}(x) + \nablar_\nu\,\nablar_\lambda\,\tilde{C}_{\lambda\mu}(x) }$  &  $-I_\munu$  &  $-I_\munu$  &  $-I_\munu$  &  $-I_\munu$ \\
\hline
\hline
%%%%%%%%%%%%%%%%%%%%%%%%%%%%%%%%%%%%%%%%%%%%%%%%%%%%%%%%%%%%%%%
%  partial derivative of quark-bilinears
%
% - no contributions, because we assume
%   translational invariance
%%%%%%%%%%%%%%%%%%%%%%%%%%%%%%%%%%%%%%%%%%%%%%%%%%%%%%%%%%%%%%%
%  l_mu g_nu
%%%%%%%%%%%%%%%%%%%%%%%%%%%%%%%%%%%%%%%%%%%%%%%%%%%%%%%%%%%%%%%
      $ \chibar\,(\tau^3)^{m}\,\rbrackets{ \nablal_\mu\,\gamma_\nu\,\gammafive^{l} \pm \gammafive^{l}\,\gamma_\nu\,\nablar_\mu }\,\chi(x)  + [\mu \leftrightarrow \nu] $  &  $(-)^{l+m} I_\munu$  &  $(-)^{l} I_\munu$  &  $(-)^{l+m}I_\munu$  &  $(-)^{l}I_\munu$ \\
\end{longtable}
}  % of fontsize
\end{center}

%%%%%%%%%%%%%%%%%%%%%%%%%%%%%%%%%%%%%%%%%%%%%%%%%%%%%%%%%%%%%%%%%%%%%%%%%
% derivative_operator_symtranstab_partII.tex
%
% Sun Nov 23 19:03:15 EET 2014
%
% PURPOSE
%%%%%%%%%%%%%%%%%%%%%%%%%%%%%%%%%%%%%%%%%%%%%%%%%%%%%%%%%%%%%%%%%%%%%%%%%
\begin{center}
  {\footnotesize
  \begin{longtable}{c|r|r|r}
     operator  &  $\mathcal{C}$  &  $\mathcal{P}\,\mathcal{D}\,[-m]\,[-r]$  &  $\ronetwofive\,\mathcal{D}\,[-\mu]$ \\
\hline
$r^k m^{n_m} \mu^{n_{\mu}}$  &  +1  &  $(-)^{n_m}(-)^{k}$  &  $(-)^{n_\mu}$ \\
\hline
%%%%%%%%%%%%%%%%%%%%%%%%%%%%%%%%%%%%%%%%%%%%%%%%%%%%%%%%%%%%%%%
% r_mu r_kappa C_kappa_nu
%%%%%%%%%%%%%%%%%%%%%%%%%%%%%%%%%%%%%%%%%%%%%%%%%%%%%%%%%%%%%%%
      $ \Tr{ \nablar_\mu\,\nablar_\lambda\,C_{\lambda\nu}(x) + \nablar_\nu\,\nablar_\lambda\,C_{\lambda\mu}(x) }$  &  $+1$  &   $I_\munu$   &   $+1$ \\
      $ \Tr{ \nablar_\mu\,\nablar_\lambda\,\tilde{C}_{\lambda\nu}(x) + \nablar_\nu\,\nablar_\lambda\,\tilde{C}_{\lambda\mu}(x) }$  &  $N$  &   $-I_\munu$   &   +1 \\
\hline
\hline
%%%%%%%%%%%%%%%%%%%%%%%%%%%%%%%%%%%%%%%%%%%%%%%%%%%%%%%%%%%%%%%
%  partial derivative of quark-bilinears
%
% - no contributions, because we assume
%   translational invariance
%%%%%%%%%%%%%%%%%%%%%%%%%%%%%%%%%%%%%%%%%%%%%%%%%%%%%%%%%%%%%%%
%  l_mu g_nu
%%%%%%%%%%%%%%%%%%%%%%%%%%%%%%%%%%%%%%%%%%%%%%%%%%%%%%%%%%%%%%%
      $ \chibar\,(\tau^3)^{m}\,\rbrackets{ \nablal_\mu\,\gamma_\nu\,\gammafive^{l} \pm \gammafive^{l}\,\gamma_\nu\,\nablar_\mu }\,\chi(x)  + [\mu \leftrightarrow \nu] $  &  $\pm 1$   &  $(-)^{l}I_\munu$  &  $(-)^{m}$ \\
\end{longtable}
}
\end{center}

\section{Symmetry properties of $S_7$}

\label{app:sseven}

In \tab{tab:ssevensym} we list all possible terms of mass dimension 7 appearing in an expansion of the effective action to order $a^3$. We
discuss their transformation properties under the $\ronetwofive$ and $\mathcal{P} \times \sbrackets{\mu_q \to -\mu_q}$ symmetries which are symmetries
of the continuum twisted mass action.  We restrict the discussion to operators involving the twisted mass $\mu_q$ only since the bare quark mass
$m_q=0$ at maximal twist. We note further that neither $\ronetwofive$ nor $\mathcal{P} \times \sbrackets{\mu_q \to -\mu_q}$ is affected by
commuting two different derivative operators in a given expression such that we omit the commuted expressions. $G_{\mu \nu}$ and $\tilde{G}_{\mu \nu}$
denote the continuum field strength tensor and its dual, respectively.

In the four fermion operators we have included a generic transformation matrix $T^A = \tau^\mu \times t^a \times \Gamma$ 
where $\tau\in\{\tau^0, \tauone, \tautwo, \tauthree\}$, 
$\Gamma \in \{ \mathbb{1}, \gamma_\mu, \sigma_{\mu\nu}, \gammafive \gamma_\mu, \gammafive \}$ and $t^a$ are acting in flavor-, Dirac- and
color-space,
respectively. Their index $A$ used as a short-hand notation for {flavor-,} Dirac- and color-indices is summed over in the fermion bilinear
product. Different Dirac structures are related via Fierz-identities and have the same transformation properties under the symmetries. Since $T^A$ is
appearing twice in all products this introduces an even number of both flavor- and Dirac-matrices such that the symmetry transformation is the same as
for the trivial product with all matrices equal to the identity.

\begin{table}[h]

\begin{center}   

     \begin{tabular}{c|c|c||c|c|c}

       operator 														& $\ronetwofive$ & $\mathcal{P}\,[-\mu_q]$&

       operator 														& $\ronetwofive$ & $\mathcal{P}\,[-\mu_q]$\\

       \hline

       $\mu_q^4 \bar \chi \chi$   												& -1& +1&

       $\mu_q^4 \bar \chi \gammafive \tauthree \chi$   										& +1& -1\\

       \hline

       $\mu_q^3 \bar \chi \slashed{D} \chi$   											& +1& -1&          

       $\mu_q^3 \bar \chi \gammafive \tauthree \slashed{D} \chi$   								& -1& +1\\

       $\mu_q^3 \tr{G_{\mu\nu} G_{\mu\nu}}$   											& +1& -1&-&-&-\\      

       \hline

       $\mu_q^2 \bar \chi D^2 \chi$   												& -1& +1&

       $\mu_q^2 \bar \chi \gammafive \tauthree D^2 \chi$   									& +1& -1\\       

       $\mu_q^2 \bar \chi \sigma_{\mu\nu} G_{\mu\nu}\chi$   									& -1& +1&

       $\mu_q^2 \bar \chi \gammafive \tauthree \sigma_{\mu\nu} G_{\mu\nu}\chi$   						& +1& -1\\       

       \hline       

       $\mu_q (\bar \chi T^A \chi)^2$												& +1& -1& 

       $\mu_q (\bar \chi \gammafive \tauthree  T^A\chi) \; (\bar \chi  T^A\chi)$   						& -1& +1\\       

       $\mu_q (\bar \chi \gammafive \tauthree  T^A\chi)^2$   									& +1& -1& -& -&-\\

       &&&&&\\

       $\mu_q \bar \chi \slashed{D} \sigma_{\mu\nu} G_{\mu\nu}\chi$   								& +1& -1&       

       $\mu_q \bar \chi \gammafive \tauthree  \slashed{D} \sigma_{\mu\nu} G_{\mu\nu}\chi$   					& -1& +1\\

       $\mu_q \bar \chi \slashed{D} D^2\chi$   											& +1& -1&       

       $\mu_q \bar \chi \gammafive \tauthree  \slashed{D} D^2 \chi$   								& -1& +1\\

       $\mu_q \bar \chi \gamma_\mu D^3_\mu \chi$   										& +1& -1&       

       $\mu_q \bar \chi \gammafive \tauthree  \gamma_\mu D^3_\mu \chi$   							& -1& +1\\

       $\mu_q \bar \chi \gamma_\mu [D_\nu, G_{\mu\nu} ] \chi$   								& +1& -1&       

       $\mu_q \bar \chi \gammafive \tauthree  \gamma_\mu [D_\nu, G_{\mu\nu} ]\chi$   						& -1& +1\\       

       \hline       

       $(\bar \chi  T^A \chi)(\bar \chi \slashed{D}  T^A \chi)$   								& -1& +1&

       $(\bar \chi \gammafive \tauthree  T^A \chi)(\bar \chi \slashed{D}  T^A \chi)$						& +1& -1\\

       $(\bar \chi \gammafive \tauthree  T^A \chi)(\bar \chi \gammafive \tauthree  T^A \slashed{D} \chi)$ 			& -1& +1& -& -& -\\

       &&&&&\\

       $\bar \chi \slashed{D} \gamma_\mu D^3_\mu \chi$   									& -1& +1&       

       $\bar \chi \gammafive \tauthree  \slashed{D} \gamma_\mu D^3_\mu \chi$   							& +1& -1\\       

       $\bar \chi  \slashed{D} \gamma_\mu[D_\nu, G_{\mu\nu} ] \chi$   								& -1& +1&       

       $\bar \chi \gammafive \tauthree   \slashed{D} \gamma_\mu [D_\nu, G_{\mu\nu} ]\chi$ 					& +1& -1\\

       $\bar \chi D^2 \sigma_{\mu\nu} G_{\mu\nu}\chi$   									& -1& +1&

       $\bar \chi \gammafive \tauthree D^2 \sigma_{\mu\nu} G_{\mu\nu}\chi$   							& +1& -1\\  

       $\bar \chi \sigma_{\kappa\lambda} G_{\kappa\lambda} \sigma_{\mu\nu} G_{\mu\nu}\chi$   					& -1& +1&

       $\bar \chi \gammafive \tauthree \sigma_{\kappa\lambda} G_{\kappa\lambda} \sigma_{\mu\nu} G_{\mu\nu}\chi$ 		& +1& -1\\       

       $\bar \chi ( D^2 )^2 \chi$   												& -1& +1&

       $\bar \chi \gammafive \tauthree  (D^2)^2 \chi$ 		  								& +1& -1\\

       $\bar \chi D^4  \chi$   													& -1& +1&

       $\bar \chi \gammafive \tauthree  D^4 \chi$ 		  								& +1& -1\\       

       $\bar \chi \gammafive G_{\mu\nu} \tilde{G}_{\mu\nu} \chi$  								& -1& +1&       

       $\bar \chi \tauthree  G_{\mu\nu} \tilde{G}_{\mu\nu} \chi$ 	  							& +1& -1\\

       &&&&&\\

       $\bar \chi G_{\mu\nu} G_{\mu\nu} \chi$  											& -1& +1&       

       $\bar \chi \gammafive  \tauthree  G_{\mu\nu} G_{\mu\nu} \chi$ 								& +1& -1\\       

       $\bar \chi \chi \tr{G_{\mu\nu} G_{\mu\nu}}$  										& -1& +1&       

       $\bar \chi \gammafive \tauthree \chi \tr{G_{\mu\nu} G_{\mu\nu}}$ 	  						& +1& -1\\

       $\bar \chi G_{\mu\nu} \tilde{G}_{\mu\nu} \chi$  										& +1& -1&       

       $\bar \chi \gammafive  \tauthree  G_{\mu\nu} \tilde{G}_{\mu\nu} \chi$ 							& -1& +1\\       

       $\bar \chi \chi \tr{G_{\mu\nu} \tilde{G}_{\mu\nu}}$  									& +1& -1&       

       $\bar \chi \gammafive \tauthree \chi \tr{G_{\mu\nu} \tilde{G}_{\mu\nu}}$   						& -1& +1\\       

\hline
\hline
     \end{tabular}

     \label{tab:ssevensym}
\caption{Transformation properties of operators appearing in $S_7$.}

\end{center}     

\end{table}

\clearpage

\bibliographystyle{JHEP}
\bibliography{improvement}

\end{document}